\documentclass[sigconf, nonacm]{acmart}
\usepackage{makecell}
\usepackage{longtable} 
\usepackage{booktabs}  
\usepackage{listings}
\usepackage{multirow}
\usepackage{diagbox}
\usepackage{float}
\usepackage{enumitem}
\floatstyle{plain}
\newfloat{CodeListing}{htbp}{lop}
\floatname{CodeListing}{Listing}

\usepackage{lmodern}
\usepackage[T1]{fontenc} 
\usepackage[utf8]{inputenc} 
\usepackage{ifthen}
\usepackage{graphicx}
\usepackage{hyperref}
\usepackage{multirow}
\usepackage{array} 
\usepackage{arydshln} 
\usepackage[many]{tcolorbox} 

\usepackage{xspace}
\usepackage{xcolor}
\usepackage[normalem]{ulem} 			
\usepackage{amsmath,amsfonts}
\usepackage{amssymb}

\usepackage{caption} 
\usepackage{subcaption} 

\newcolumntype{L}[1]{>{\raggedright\let\newline\\\arraybackslash\hspace{0pt}}p{#1}}
\newcolumntype{C}[1]{>{\centering\let\newline\\\arraybackslash\hspace{0pt}}p{#1}}
\newcolumntype{R}[1]{>{\raggedleft\let\newline\\\arraybackslash\hspace{0pt}}p{#1}}

\newboolean{showcomments}
\setboolean{showcomments}{true}

\ifthenelse{\boolean{showcomments}}
{
	\newcommand{\nb}[3]{
		{\colorbox{#2}{\bfseries\sffamily\scriptsize\textcolor{white}{#1}}}
		{\textcolor{#2}{\sf\small$\blacktriangleright$\textit{#3}$\blacktriangleleft$}}}
	 
	\newcommand{\bnote}[2]{\fbox{\color{blue}\bfseries\sffamily\scriptsize#1}
    	{\color{blue}\sf\small$\blacktriangleright$\textit{#2}$\blacktriangleleft$}}
	\newcommand{\old}[1]{{\color{gray}\sout{#1}}} 
	\newcommand{\del}[1]{\old{#1}} 
	\newcommand{\ins}[1]{{\textcolor{blue}{\uline{#1}}}} 
	\newcommand{\ugh}[1]{{\textcolor{red}{\uwave{#1}}}} 
	\newcommand{\chg}[2]{{\textcolor{red}{\sout{#1}}}{\ra}\textcolor{blue}{\uline{#2}}} 
	 
	\newcommand{\fix}[1]{\bnote{FIX}{#1}}
	\newcommand{\sd}[1]{\bnote{SD:}{#1}}
}{
	\newcommand{\bnote}[2]{}
	\newcommand{\nb}[3]{}
	\newcommand{\old}[1]{}
	\newcommand{\del}[1]{}
	\newcommand{\ins}[1]{}
	\newcommand{\ugh}[1]{}
	\newcommand{\chg}[2]{}
	
	\newcommand{\sd}[1]{}
	\newcommand{\fix}[1]{}
} 

\newcommand{\hide}[1]{}

\usepackage[many]{tcolorbox}  

\graphicspath{{figures/}{newFigures/}}

\newcommand{\commented}[1]{}

\newcommand{\eg}{\emph{e.g.,}\xspace}
\newcommand{\ie}{\emph{i.e.,}\xspace}
\newcommand{\etal}{\emph{et al.,}\xspace}

\usepackage{url}            
\makeatletter
\def\url@leostyle{%
  \@ifundefined{selectfont}{\def\UrlFont{\sf}}{\def\UrlFont{\small\sffamily}}}
\makeatother
\definecolor{main}{HTML}{828282}    
\definecolor{sub}{HTML}{E0E0E0}     

\tcbset{
    sharp corners,
    colback = white,
    before skip = 0.2cm,    
    after skip = 0.5cm      
}                           

\newtcolorbox{cbox}{
    enhanced, 
    boxrule = 0pt, 
    borderline = {0.75pt}{0pt}{main}, 
    borderline = {0.75pt}{2pt}{sub} 
}

\newcommand{\esope}{Esope\xspace}
\newcommand{\seg}{segment\xspace}
\newcommand{\segs}{segments\xspace}
\newcommand{\Seg}{Segment\xspace}
\newcommand{\Segs}{Segments\xspace}
\newcommand{\fold}{FORTRAN~77\xspace}
\newcommand{\fnew}{Fortran~2008\xspace}

\definecolor{verylightgray}{gray}{0.93}
\newcommand{\trimcolorbox}{\vrule width 0pt height .99\ht\strutbox depth .99\dp\strutbox\relax}
\newcommand{\mycode}[1]{%
  \begingroup
  \setlength{\fboxsep}{0pt}%
  \colorbox{verylightgray}{\trimcolorbox\texttt{#1}\/}%
  \endgroup
}

\begin{document}

\pagestyle{plain}

\title{
  Migrating \esope to \fnew using model transformations
}

\author{Younoussa Sow}
\orcid{0000-0000-0000-0001}
\affiliation{%
 \institution{DTIPD Framatome \\ Univ. Lille - Inria - CNRS - Centrale Lille - UMR 9189 CRIStAL} 
 \city{Villeneuve d'Ascq}
 \country{France}
}
\email{younoussa.sow@framatome.com} 

\author{Nicolas Anquetil}
\orcid{0000-0000-0000-0001}
\affiliation{%
 \institution{Univ. Lille - Inria - CNRS - Centrale Lille - UMR 9189 CRIStAL}
 \city{Villeneuve d'Ascq}
 \country{France}
}
\email{nicolas.anquetil@inria.fr} 

\author{Léandre Brault}
\orcid{0000-0000-0000-0001}
\affiliation{%
 \institution{DTIPD Framatome}
 \city{Paris}
 \country{France}
}
\email{leandre.brault@framatome.com}

\author{St\'ephane Ducasse}
\orcid{0000-0000-0000-0001}
\affiliation{%
 \institution{Univ. Lille - Inria - CNRS - Centrale Lille - UMR 9189 CRIStAL}
 \city{Villeneuve d'Ascq}
 \country{France}
}
\email{stephane.ducasse@inria.fr}

\begin{abstract}
Legacy programming languages such as \fold still play a vital role in many industrial applications. 
Maintaining and modernizing these languages is challenging, especially when migrating to newer standards such as \fnew.
This is exacerbated in the presence of legacy proprietary extensions on such legacy languages, because their semantics are often 
based on old context (limits of legacy language, domain logic,...).
This paper presents an approach for automatically migrating  \fold with a proprietary extension, named \esope,  to \fnew.
We introduce a tool that converts \esope source code to \fnew.
While supporting readability of the generated code, we want to maintain the level of abstraction provided by Esope.
Our method uses model-driven engineering techniques, with transformations to generate a target model 
from which we export easy-to-read \fnew source code.
We discuss the advantages, limitations, and maintainability considerations of our approach and provide insights into its scalability and adaptability to evolving requirements.
\end{abstract}

\maketitle

\section{Introduction}

Many organisations continue to rely on legacy systems written in outdated programming languages such as \fold~\cite{Vand22a}, 
sometimes supplemented by proprietary constructs to compensate for the lack of functionality of these old languages~\cite{Tine09a}.
These systems remain essential to the business, but are under increasing pressure to adapt to modern platforms and development practices.
We are experiencing these difficulties as we were called,  by the Framatome company\footnote{\url{https://www.framatome.com/en/} Framatome designs, builds, and services nuclear steam supply systems.}, on a project to migrate \fold with \esope (a proprietary extension) to \fnew.
\esope was created by the CEA\footnote{\url{https://www.cea.fr/english}} at the early eighties, as an extension to \fold.
Its goals were to facilitate the management of data by allowing dynamic allocation of data structures.
However, the evolution of Fortran compilers made some of the assumptions on which the design of \esope was based unreliable.
As a consequence, developers have to reduce the level of optimisation during compilation to get a working executable.
While this solution works today, it raises the question of whether these systems will be compilable in the long term.
Migrating these systems to modern languages such as \fnew, which natively offers the concepts provided by \esope, can guarantee the longevity, compatibility, and maintainability of these systems.

\esope is actually a transpiler to \fold, adding instructions to manage its own heap memory.
As such, the code it generates is low level and would not be a good base for migration to \fnew.
Future maintainers of the code should not have to deal with this low level memory management but focus on the domain abstractions currently expressed in \esope.

This article focuses on the migration of \fold with \esope to \fnew. 
More specifically, it addresses the challenges of transforming an \esope Abstract Syntax Tree (AST) representation into a \fnew target AST.
Our approach mixes two techniques commonly used in model-to-text transformations (visitors and string templates) to handle the migration, offering flexibility and modularity.
It allows us to handle scenarios where instructions in \esope translate into much larger pieces of \fnew with minimal variations (in some cases, expanding by a factor of a hundred).

The contributions of this article are: 
\begin{enumerate}[label=\textbf{(\arabic*)},wide]
  \item Proposes an implementation in \fnew of structured data types that can be enlarged.

  \item Handles these structured types the ``\esope way'' that is to say, freeing the developers from having to handle possible pointer changes after enlarging a structure.

  \item Positions our approach in the landscape of model transformation tools.

  \item Proposes a transformation technique that makes it easier to generate large chunks of similar text.

\end{enumerate}

In the remaining of the paper, we will highlight the technical and conceptual challenges of such a transformation, 
as well as the adopted methodology to address them. 

Section~\ref{sec:source-code-transformation} provides an overview of past Fortran migration solutions and introduces several model transformation approaches.
Section~\ref{sec:context-of-the-migration} outlines the specific context of this project by presenting the \esope language and detailing the motivations behind the migration and the constraints involved.
Section~\ref{sec:from-esope-to-fnew} motivates the choice of \fnew as the target language and examines the migration's expectations by detailing the problems faced with \esope, Fortran, and the company's requirements (\eg readability and maintainability of the translated code).
Section~\ref{sec:segment-migration} addresses these challenges by presenting our design and architectural approach.
Section~\ref{sec:migration-implementation} describes the practical implementation of these solutions.
Section~\ref{sec:future-work} reflects on the achievements of the migration, identifies current limitations, and suggests directions for future work.
Finally, we summarize the project's main contributions and lessons learned.

\section{Source Code Transformation} \label{sec:source-code-transformation}

In this section  we begin with a discussion of past Fortran migration solutions.
Then, we review existing transformation approaches, focusing on model driven engineering.

\subsection{Fortran Migration}  \label{sec:sota-fortran-migration}

Several studies were conducted on the migration of legacy Fortran code to other languages including:
Legacy Fortran to Python~\cite{Bysi16a}, JVM Bytecode or Java~\cite{Fox97a,Seym01a,Dool99a},
C/C++~\cite{Feld95a,Feld91a,Gros12a}, Pascal~\cite{Frea81a}, Basic~\cite{Cari92a}, 
Ada~\cite{Slap83a,Pars88a}, Algol~\cite{Prud77a}.

An issue with those solutions is that the source and target languages differ, most of them treated code migration from Fortran to some other language.
This means that they are not constrained by the need to maintain familiarity with the transformed code, and thus need not attempt to generate familiar code to the developers.

Fieldman \etal~\cite{Feld95a,Feld91a} carried out the first attempt to automatically migrate from Fortran to C language.
This work led to the development of the f2c tool, which automatically converts Fortran code to C.
The tool is used as a front-end compiler in conjunction with a traditional C compiler.

Mariano \etal in~\cite{Mari19a}, propose an approach to modernize legacy \fold to more modern Fortran (Fortran 90 or 95).
But this approach is manual.

All these automatic migration tools generate compilable code, with little consideration given to readability. 
This makes the generated code difficult to maintain and evolve.

We note that, modernizing legacy Fortran involves dealing with ``outdated'' language constructs.
For example, the elimination of unstructured code such as 
\mycode{GOTO} statement~\cite{Esos94a}, \mycode{Arithmetic-If}~\cite{Cecc14a},
\mycode{COMMON} blocks~\cite{Dool99a, Orch13a}, \mycode{EQUIVALENCE}~\cite{Gros12a}.

\subsection{Model Transformation}

Model transformation is a cornerstone of model-driven engineering (MDE) or model-driven development (MDD)~\cite{Kolo08a}, 
enabling the automation of tasks such as code generation, code refactoring, and code migration.
Model transformations are categorized into two primary paradigms~\cite{Kaha19a}:
\emph{Model-to-Model} (M2M): These transformations operate between models at similar levels of abstraction.
  They can adhere to the same metamodel (\eg migrating from Java~8 to Java~17) or adhere to different metamodels (\eg migrating from Fortran to Java);
\emph{Model-to-Text} (M2T): These generate textual artifacts (\eg source code, documentation) from models, 
  bridging abstract representations with concrete implementations.

There are three main \emph{Model-to-Model} transformation approaches:
\begin{itemize}[wide]
  \item \textbf{Rule-based approaches}: Tools such as ATL~\cite{Joua06a} and QVT~\cite{Witk05a} map models declaratively and offer high traceability. 
  However, scalability can be a challenge for complex models.
  If it is easy to define a query finding a specific entity in a small model, it becomes harder to do it for a large one;

  \item \textbf{Operational approaches}: Imperative tools such as ATL, QVTo, and Xtend~\cite{Hebi18a} provide fine-grained control, 
  which is ideal for handling complex mappings, but requires significant manual effort. One needs to define each operation to apply to each node of the AST;

  \item \textbf{Graph-based approaches}: Tools such as Henshin~\cite{Aren10a} use graph rewriting rules for intuitive and 
  visual transformation, but suffer from performance issues with large models.
\end{itemize}

For \emph{Model-to-Text} transformation, there are two main approaches:
\begin{itemize}[wide]
  \item \textbf{Template-based approaches}: Tools such as Acceleo~\cite{Nate23a} use predefined templates to generate text, 
  ensuring readability but limiting flexibility for complex structures;

  \item \textbf{Visitor-based approaches}: Tools such as Xtend~\cite{Boro18a} traverse model structures programmatically, 
  allowing fine-tuned output generation but increasing development overhead.
\end{itemize}

This paper innovates by hybridising \emph{Model-to-Text} (visitor and string template) for a \emph{Model-to-Model} transformation of Esope ASTs.

\section{Context of the migration} \label{sec:context-of-the-migration}

As a global leader in nuclear energy, Framatome provides innovative solutions for designing,
constructing, and operating nuclear power plants. The company specializes in nuclear reactor fuel,
services, and instrumentation with a strong focus on safety and sustainability. 
Headquartered in France, it operates in over 30 countries.

In this section we give the industrial context of this project, presenting \esope, the proprietary programming language that must be migrated.
We emphasize some important points that impact the migration and finish by explaining why this project is necessary.

\subsection{A brief overview of \esope} \label{sec:esope-introduction}

\fold programmers are confronted with two primary challenges: (i) the language does not provide a record type construction (like a \mycode{struct} in C), and (ii) it does not offer either any dynamic memory management solution.

In order to address these issues, \esope \cite{Bert20a} was created as an extension to \fold in the early eighties.
The goal was to create a programming language allowing (1) complex data structures, (2) automatic memory management, and (3) memory swap management.

This is how an entity called \emph{\seg} was added, as well as primitives for manipulating it.
A \seg is a group of Fortran variables, called \emph{fields}, that are defined by the programmer.
The field definitions have the same syntax as regular variable definitions in Fortran.
An instance of a \seg is referenced by a variable called a \emph{pointer}.
Knowing the pointer is enough to access all the fields contained in the segment.

Listing~\ref{lst:esope-example} is an example of \esope code for the management of users inside a book library:
\begin{itemize}[wide]
  \item Lines 3 to 6 declare a segment for a \mycode{user} with a \mycode{uname} ``user name'' (line 4) and an array 
  of borrowed books (User Borrowed Books, \mycode{ubb}, line 5).
  The size of the array is dynamically defined by the variable \mycode{ubbcnt};

  \item Line 7, a new statement, \mycode{pointeur}\footnote{french for ``pointer''}, was introduced to define a pointer variable (\mycode{ur}) pointing to a \mycode{user} segment;

  \item Line 8, is a \fold comment: marked with a `C' (or a `*') as first character of the line;

  \item Line 10, gives an example of one of the new statements introduced to manipulate segments.
    Here, \mycode{segini} creates an ``instance'' of \mycode{user} and \mycode{ur} will point to it;

  \item Line 11, a new dotted notation allows one to access fields of a segment, here to set the \mycode{name} of the user;
	
  \item Line 12, another new notation (slash notation) allows one to get the size of the first (``\mycode{/1}'') dimension of array \mycode{ur.ubb}.
  Here it will return the value of \mycode{ubbcnt} when the segment was created.
\end{itemize}

\begin{CodeListing}
\begin{lstlisting}[language=Fortran, label=lst:esope-example, caption=Example of \esope code (see text for explanations)]
      SUBROUTINE NEWUSER(LIB,NAME)
      INTEGER UBBCNT
      SEGMENT, USER
       CHARACTER*40 UNAME
       INTEGER UBB(UBBCNT)
      END SEGMENT
      POINTEUR UR.USER
C the user does not have a book yet 
      UBBCNT = 0
      SEGINI, UR
      UR.UNAME = NAME
      WRITE(*,*) UR.UBB(/1)
[...]
\end{lstlisting}
\end{CodeListing}

\subsection{\Seg type} \label{sec:segmentdesc}

Each segment is instantiated inside some routine through a pointer, that supports passing segments as arguments to subprograms.
Note that in this paper, we call segment the structured type that contains fields as well as instances of this type that are pointed to by pointer variables.

\Segs can contain other segments to create structures such as graphs.
\Segs may also contain arrays.
In \fold, the size of arrays must be known at compile time.
\esope introduced the ability to delay the sizing of arrays at run time with the concept of \textit{dimensioning variable} and \textit{dimensioning expression}.
This feature is only available to arrays inside segments.
Listing~\ref{lst:esope-example}, line 5 shows an example of this where the \textit{dimensioning variable} is \mycode{ubbcnt} and the \textit{dimensioning expression} is simply the variable, but could be more complex (eg. \mycode{ubbcnt*1.1} to give 10\% extra space in the array).

The \textit{dimensioning variable} must be assigned before the segment is allocated (see Section~\ref{sec:segcmd}).

Note that as an added complexity, Fortran does not impose to declare a variable (implicit typing), so line 2 of Listing~\ref{lst:esope-example} is optional.

Dimensioning arrays at run time offers the possibility to resize arrays and hence their segment at run time.
This is equivalent to the \mycode{realloc} function in C language.
In C, \mycode{realloc} takes an address as an argument and the new size.
It returns a (possibly new) address for the re-dimensioned structure.
\esope offers a similar feature, where the new size is given by a new value of the \textit{dimensioning variable}.
But, differently to C, the pointer to the structure is guaranteed to remain the same.
\esope achieves this by managing a ``double indirection'':
The ``pointer'' handled by the developer is actually an index in a large array of all the ``descriptors'' of existing segment instances.
When re-dimensioning a \seg, the content of its descriptor, such as the memory address of its fields might change, but the index of its descriptor, known to  the developer as a ``pointer'', never changes.

Finally, for each \seg, \esope automatically declares a default pointer with the same name as the segment, as if the \mycode{POINTEUR user.user} statement had been given after the segment definition.
Along with that, fields of a segment can be manipulated without the default pointer prefix.
For example, using the default pointer, the statement on line 11 of Listing~\ref{lst:esope-example}, could become \mycode{uname = name} where \mycode{uname} means the field  \mycode{uname} in the segment pointed to by the default pointer \mycode{user}.

\subsection{\Seg manipulation} \label{sec:segcmd}

The following \esope statements  operate on pointer variables:

\begin{description}[wide]
  \item[Declaring a pointer to a \seg] (\mycode{pointeur p.user}) \\ Declares a variable \mycode{p} pointing to an instance of \mycode{user}.

  \item[Allocating \seg] (\mycode{segini p}).
  It allocates memory for an instance of a \seg and put its address in \mycode{p}.

  \item[Freeing \seg] (\mycode{segsup p})
  It frees the memory pointed to by \mycode{p}.
  The pointer variable is set to zero ($0$) which is the null pointer.
  
  \item[Resizing \seg] (\mycode{segadj p})
  The array fields, with \textit{dimensioning variable}, of an existing instance of a \seg (pointed by \mycode{p})
  are resized (enlarged or shrinked). The instance (value of the pointer) itself remains unchanged.
  This statement preserves existing data in the instance.

  \item[Printing \seg] (\mycode{segprt p}) This is used to display the name, dimensions and value of each field of a \seg.
  Each field is printed on a separate line;
  
  \item[Swapping-out \seg] (\mycode{segdes p}) This indicates that the \seg can be moved to ``swap'' in the event of a memory shortage.
  It also marks the pointed \seg as ``inactive'';
  
  \item[Swapping-in \seg] (\mycode{segact p}) If the \seg was swapped out, this indicates that it needs to be moved back into the main memory.
  It also marks the \seg as ``active'';

  \item[Duplicating a \seg] \label{segcmd-segcop}
  (\mycode{segini newp=oldp}) 
  This is an alternate use of the \mycode{segini} statement where a new instance of the segment is allocated, its address put in the pointer \mycode{newp}, and the content of the existing instance (\mycode{oldp}) is copied to the new one.
  The old instance is kept.
  The state of \mycode{oldu} is not modified.
  Both \mycode{newu} and \mycode{oldu} must be instance of the same \seg.
  This is informally referred to as a ``segcop'' by the developers.

  \item[Copying a \seg] \label{segcmd-segmov} 
  (\mycode{segact newp=oldp})
  This is an alternate use of the \mycode{segact} statement, where an instance of a \seg from the source (\mycode{oldp}) is copied to the destination (\mycode{newp}). 
  The target segment must already exist (ie. having been allocated) before this statement.
  The statement also performs a \mycode{segact} on \mycode{newp} beforehand (in case the \seg was swapped out).
  This is informally referred to as a ``segmov'' by the developers.

\end{description}

\subsection{Migration motivation} \label{sec:reasonOfMigration}

\esope is a transpiler that converts \esope code into pure \fold code.
It manages its own heap memory as a huge \fold array\footnote{The heap memory array is a C array created once using \mycode{malloc}.} and converts the \esope commands as offsets in this memory zone.

A major issue with \esope is that the latest versions of the Fortran compilers are not compatible with these advanced memory management ``tricks''.
The generated code compiles, but it can no longer be highly optimized by the compiler, which is a significant drawback for High Performance Computing applications.

This is seen as raising significant doubt on the ability of future compilers to just compile the code.
As such, it was decided to migrate \esope code to modern Fortran.

\section{From Esope to \fnew} \label{sec:from-esope-to-fnew}

We just saw that our industrial partner wishes to migrate its \esope code base to modern Fortran.
We now justify the choice of \fnew as the target for the migration; then we discuss the various problems that this migration raises whether they originate in \esope itself, \fold, or requirements that the company set.

\subsection{\fnew}\label{sec:fnew}

\fnew was chosen as the target for the migration for the following reasons:

\begin{itemize}[wide]

\item Terekhov and Verhoef \cite{Tere00a} warn about the danger of subtle differences between primitive data type representations in two different languages.
By using a more modern version of the same language (Fortran), we keep the same compiler and should therefore be immune to this kind of problems.

\item \esope programs actually contain a lot of \fold code, ``peppered'' with a few \esope commands.
Remaining in the same language allows us to not translate a large part of the original code and rely on the backward compatibility of the compiler to accept it.

\item The development team, comprised solely of physicists, has extensive experience with \fold and is familiar with its  ecosystem, tools, and best practices. 
By continuing to use Fortran, we minimize the impact on their working habits.

\item Object-oriented programming support exists in \fnew: declaration of derived types and structures that seem a good match for the \segs.
It may prove useful to handle collections of heterogeneous segment types.
  
\item \fnew offers dynamic memory management that the \esope programs need.

\end{itemize}

\subsection{\esope problems}  \label{sec:pb-esope}

We envisioned the following problems for the migration of \esope concepts to \fnew:

\begin{enumerate}[label=\textbf{(\arabic*)}, wide]

  \item \label{reqt-esope-commands}
  All memory management commands (see Section~\ref{sec:segcmd}) must be re-implemented and their exact behavior preserved.
  In \esope, these commands rely on a meta-description of the \seg on which they are working.
  We must find a way to have a similar behavior, but \fnew does not offer a meta-description of its user-defined types.

  \item \label{reqt-esope-obsolete}
  Some of these commands (\mycode{segact} and \mycode{segdes}) will be obsolete because it is no longer necessary to handle swapping explicitly.
  
  \item \label{reqt-esope-double-indirection}
  As explained in Section \ref{sec:segmentdesc}, memory reallocation is handled automatically by \esope with a ``double indirection'' mechanism to free the developers from address changes when a segment is resized.
    We will need to re-implement this mechanism to avoid profound modifications of the code or disrupting developers' habits.
  
  \item \label{reqt-esope-implicit-none} 
  \fold can declare a variable implicitly based on some naming conventions.
  This is considered bad practice and the \mycode{implicit none} statement (not standard in \fold) forces developers to declare all symbols' type (variables and functions, see also point \ref{reqt-fortran-externalfct-return-type}) before they can be referenced.
  However, \esope transpiler does not recognize this statement, and it was therefore ignored in all programs.
  As a result, some variables were not explicitly declared in the \esope code.
  Framatome wishes to enforce the \mycode{implicit none} convention in all migrated programs and the migration will need to declare all variables.
  
  \item \label{reqt-esope-type-inference}
  \esope pointers are implemented as Fortran integers and sometimes declared as such.
	In \fnew, the pointer type does exist and is different from the integer type.
  
   \item \label{reqt-esope-negative-pointer}
   Taking advantage of the equivalence between pointer and integer, \esope offers a feature that allows a ``pointer'' to be negative, providing additional functionalities depending on the context (\eg used to declare read-only \segs).

\item \esope offers the possibility to access the fields of a segment in a simplified way for the default pointer of the segment (see Section \ref{sec:segmentdesc}).
This feature does not exist in \fnew and we will need to create the appropriate explicit accesses.
\end{enumerate}

\subsection{Fortran problems}  \label{sec:pb-fortran}

Although the migration will target the Fortran language, there are special practices and differences between \fold and \fnew that must be considered:

\begin{enumerate}[label=\textbf{(\arabic*)}, resume, wide]

  \item \label{reqt-fortran-preprocessor}
  There are preprocessor statements in the source code (\mycode{\#define}, \mycode{\#ifdef}, \dots) that Framatome wishes to eliminate~\cite{Mede18a}.
  These statements seem to be mostly used to ensure portability of the code across various computer architectures.
  This is no longer necessary, as computer architecture has now been standardised to intel x64.

  \item \label{reqt-fortran-obsolete-constructs} 
  The ``outdated'' constructs such as \mycode{COMMON} block or \mycode{GOTO} are well-known issues in Fortran migration \cite{Dool99a, Orch13a}.

  \item \label{reqt-fortran-external-statement}
  In \fold, the interface of external functions (\ie  defined in some other source file) is declared with the statement \mycode{external}
  which only specifies the name and the return type of the function and not the number and types of parameters.
  The same statement in \fnew requires to completely specify the interface.
  
  \item \label{reqt-fortran-externalfct-return-type}
  Furthermore, in \fold when using an external function (defined is some other source file), its return type must be declared independently of the \mycode{external}  statement. 
  This can be explicit (\eg \mycode{integer afunction}) or implicit (based on naming convention of the function) as described in point~\ref{reqt-esope-implicit-none}.
  In \fnew the declaration of external functions' return type is no longer independent of the \mycode{external}  statement, but one must be careful to not migrate the explicit return type definition as if it was a variable declaration (which has the same syntax in \fold).
  
  \item \label{reqt-fortran-arrays-of-pointers}
  \fnew does not allow arrays of pointers\footnote{\url{https://fortran-lang.discourse.group/t/arrays-of-pointers/4851}}.
  This is an issue because the source code contains many of these to implement collection of heterogeneous \segs.
\end{enumerate}

\subsection{Usage concerns} \label{sec:pb-nontech}

There are other concerns that are not related to programming language features, but have to do with actual usages and desirable practices for the future:
\begin{enumerate}[label=\textbf{(\arabic*)}, resume, wide]

 \item \label{reqt-nontech-include}
 There are four include syntaxes in the existing code: preprocessor (\mycode{\#include}), \fold (\mycode{include}), and \esope (\mycode{\%inc} or \mycode{-inc}).
  Framatome wishes to simplify this by using at most one syntax or possibly removing all includes~\cite{Baba17a, Miha10a}.

  \item \label{reqt-nontech-pointer-assingment}
  With the \fnew pointer, you can use either an assignment statement (\mycode{=}) or a pointer assignment statement (\mycode{=>}): 
  \mycode{p => q} copies the address contained in pointer \mycode{q} into pointer \mycode{p}; 
  and \mycode{p = q} copies the values of the structured type pointed by \mycode{q} into the structured type pointed by \mycode{p}.
  Framatome wishes to forbid the second behavior for \segs.
    
  \item \label{reqt-nontech-dimensioninvars}
  In arrays within \segs, dimensioning variables allow one to define at run time the size of the array (see Section~\ref{sec:segmentdesc}).
 A perceived issue with \esope is that nothing differentiates these variables from ``normal'' Fortran ones.
 We would like to offer the means to better differentiate these variables and control how they are manipulated.

  \item \label{reqt-nontech-dimensioninexprs}
  To implement this ``dimensioning expressions'' behavior in \fnew, there are two solutions: allocatable variables or pointer to arrays created dynamically.
   Allocatable variables are typically recommended because Fortran can automatically resize an array if needed.
   However, for consistency, Framatome does not wish to allow this behavior, which would compete with the dedicated \mycode{segadj} mechanism.

  \item  \label{reqt-nontech-newsegment-compact}
  \esope provides a compact and natural way to create user-defined types through \segs.
  \fnew is a rather verbose language and Framatome wishes to keep a simple syntax to ensure readability of the segment definitions.

  \item  \label{reqt-nontech-familiar}
  It is desirable that the generated code looks as familiar as possible to the developers.
  Formatting, comments, or empty lines should be preserved as much as possible.

  \item  \label{reqt-nontech-traceability}
  It is desirable to have a clear traceability of the migration that lets developers understand from where a given piece of code comes.
  For example, if some code is added in a file as the result of performing an include statement, this should be clear in the migrated code (\eg by adding a comment).
 
  \item  \label{reqt-nontech-obsolete-esope}
  After migration, some statements may become useless (see Section~\ref{sec:pb-esope}).
  However, and again for traceability reasons, it must be explicit in the resulting code that the instruction was removed and why it was removed.

  \item  \label{reqt-nontech-params-intent}
  Contrary to \fnew, \fold was missing a declaration of whether parameters were input or output.
  The information is sometimes made explicit in comments but not in a way that can be automatically processed.
   Framatome wishes that this declaration be now explicit as allowed in \fnew (``parameter intent'').

\item \label{reqt-nontech-dble-indirect}
	The company has a wealth of archived \segs where the relationship between \segs (like graph of \segs) depends on the double indirection (see Section~\ref{sec:pb-esope}).
	The migrated application will need to be able to reload these archived \segs and therefore handle the double indirection in a manner compatible with \esope.

  \item \label{reqt-nontech-performence}
  Obviously, Framatome wishes to keep similar performance in terms of execution time and memory usage.
\end{enumerate}

\section{Strategies for \esope Migration} \label{sec:segment-migration}

The migration of \esope to \fnew requires a methodical approach to ensure the 
preservation of existing functionalities and behaviors. 

The requirement listed in the previous section translate to the following high level tasks:
\begin{itemize}[wide]
\item Find out how to implement the \segs.

\item Implement the notion of dimensioning variables and dimensioning expressions in \fnew.

\item Offer the possibility to resize array fields in \segs.

\item Migrate the double indirection mechanism of \esope.

\item Migrate the \esope commands in \fnew.

\item Migrate some additional \esope utilities (described in Section \ref{sec:esope-introduction}): \seg field access (``dotted notation''), or array field dimensions (``slash notation'').

\item Take care of the preprocessing statements.

\item Reintroduce explicit typing of variables in the migrated code (see Section~\ref{sec:pb-nontech}).

\item Explicitly declare the parameters' intent (see Section~\ref{sec:pb-nontech}).

\end{itemize}

The following subsections discuss these points.
We do not discuss in this paper the problem of migrating ``outdated'' constructs (see point~\ref{reqt-fortran-obsolete-constructs}).
By keeping Fortran as our target language, we may ignore them in a first solution.
Migrating these constructs has been studied elsewhere (\eg \cite{Dool99a, Gros12a, Morr97a, Orch13a}) and we plan to look at existing solutions in the future.

\subsection{\esope \seg migration} \label{sec:migrate-segment}

We identified three possibilities for \seg migration:

\begin{description}[wide]
  \item[Derived Type:] It is a special form of data type that can encapsulate other built-in types as well as other derived types. 
  It could be considered equivalent to struct in C.

  \item[Parameterized Derived Type:] It is a derived type with parameters for its array attributes.
  These parameters may specify the array size (as in \esope) or field type kind \eg integers flavours, on 1, 2, 4, or 8 bytes.

  \item[C interoperability:] \fnew has explicit C interoperability, enabling the transparent use of C functions (\eg malloc or free) as Fortran ones.
\end{description}

As described in Section~\ref{sec:fnew}, 
to minimize disruption to the developer's working habits, we want to stick to the Fortran language.
Pletzer~\etal \cite{Plet08a} warn that interoperability between languages may raise some unexpected issues.
For this reason, we will not consider the C interoperability option.

Despite their apparent similarity to \esope \segs, we will not use the Parameterized Derived Types because:
\begin{itemize}[wide]
\item Parameterized derived types can be allocated at run time to a given size, but they cannot be resized;

\item Resizing can be done by creating a new Parameterized derived type, but this would involve copying the entire \seg to a new location even when shrinking the segment.
\end{itemize}

By using derived types, we have to handle \seg resizing manually, but this gives us finer control on the operation and how it is performed (see Section~\ref{sec:migrate-commands}).

Taking back the \mycode{user} example of Section~\ref{sec:esope-introduction} (Listing~\ref{lst:esope-example}, lines 3 to 6), we convert it to the derived type in Listing~\ref{lst:f2k-example}.

\begin{CodeListing}[htbp]
\begin{lstlisting}[language=Fortran, caption={Derived Type definition in \fnew},label={lst:f2k-example}]
module user_mod
  ... ! all use statements
  implicit none
  private
  type, extends(segment) :: user
    integer, private :: ubbcnt = 0 
    character(len=40), public :: uname = ''
    integer, pointer, public :: ubb(:) => null()
  end type user
  ...
end module user_mod
\end{lstlisting}
\end{CodeListing}

The new implementation exhibits the following points:
\begin{itemize}[wide]

\item Line 1, the derived type is defined in a separate module\footnote{\href{https://fortran-lang.org/learn/best\_practices/modules\_programs/}{https://fortran-lang.org/learn/best\_practices/modules\_programs/}} along with the required functions and subroutines needed to handle it (hidden in line 10).

\item Line 5, the derived type is implemented as a subclass of the abstract \mycode{segment} derived type 
(see section~\ref{sec:migrate-commands}).

\item Line 6, the \mycode{ubbcnt} dimensioning variable is declared as a private field of the derived type.
  This will be further discussed in Section \ref{sec:migrate-dimensioning-expr}.

\item Line 7, the segment fields are declared in the derived type according to \fnew syntax, here the \mycode{uname} field.

\item Line 8, the array fields with dynamic size are declared with the ``deferred-shape'' feature of \fnew.
We manually allocate dynamically when creating the instance.
   The \mycode{pointer} attribute in the declaration will be discussed in Section \ref{sec:migrate-array-field}.

\end{itemize}

The declaration of a pointer to a ``user''  (line 7 of Listing~\ref{lst:esope-example}) will become: \mycode{type(user), pointer :: ur}.

\subsection{Dimensioning variables and expressions} \label{sec:migrate-dimensioning-expr}

In \esope dimensioning variables were plain Fortran variables (Section~\ref{sec:esope-introduction}, Listing~\ref{lst:esope-example}, line 2) declared next to the \seg declaration to ease understanding.
It was often accompanied by a comment describing its use in the \seg (see also Section \ref{sec:future-work}).

Framatome wishes to enforce better control on these variables (Section \ref{sec:pb-nontech}, point~\ref{reqt-nontech-dimensioninvars}).
For this, we propose to declare the dimensioning variable inside the derived type as a private field.
This will force the developers to modify its value only through a subroutine that we will offer.

The dimensioning expressions that give the actual size of array fields in the segments are arithmetic expressions involving one or more dimensioning variable(s).
These expressions are needed to instantiate the segment (command \mycode{segini}) or to resize it (command \mycode{segadj}).
For convenience, we chose to implement each dimensioning expression as a private function in the derived type's module.
It takes as parameters all the dimensioning variables of the segment and returns the expected value for the given array field it dimensions.

This implementation choice allows developers to easily find the dimensioning expressions and modify them when needed.
It would also allows one to put additional controls to ensure that the dimensioning variables or expressions fall within an acceptable range.
This is an additional security that did not exist in \esope but would need to be implemented by the developers on a per variable/expression basis.

\subsection{Dynamic array fields}  \label{sec:migrate-array-field}

In \fnew, there are three possibilities to declare arrays with a dimension known only at run time:

\begin{itemize}[wide]
\item dimensioning expression with value known at runtime \\ (\eg \mycode{integer :: arr(n)} where \mycode{n} is a variable).
\item pointer to an \emph{deferred-shape}\footnote{\url{https://j3-fortran.org/doc/year/24/24-007.pdf\#subsubsection.1080}} array \\ (\eg \mycode{integer,} \mycode{pointer :: arr(:)}).
\item \emph{allocatable} declaration of a deferred-shape array \\ (\eg \mycode{integer,allocatable :: arry(:)}).
\end{itemize} 

The dimensioning expression in the declaration is not possible because the dimensioning variable is part of the derived type (as a private field, Sections \ref{sec:migrate-segment} and \ref{sec:migrate-dimensioning-expr}).

The allocatable solution is a recommended good practice as it delegates memory management to the compiler.
Yet, we excluded it because it may automatically re-allocate an array each time an assignment is made to it.
Framatome does not wish this behavior and prefers explicit errors to be raised in such a case (see Section~\ref{sec:pb-nontech}, point~\ref{reqt-nontech-dimensioninexprs}).

We therefore chose the deferred-shape solution even though it forces us to implement the proper memory (de-)allocation routines.

Note that Framatome does not wish to allow the \mycode{=} operator on pointers (copy of the derived types, see Section~\ref{sec:pb-nontech},  point~\ref{reqt-nontech-pointer-assingment}) because the instances might not have the same sizes.
It wants to enforce using the \mycode{=>} (pointer assignment) operator.
Therefore we will override the \mycode{=} operator definition to raise an error at compile time.

\subsection{Graph of \segs} \label{sec:migrate-double-indirection}

In Framatome code, it is common to have a graph of segments referring one to the other. 
Such a graph needs to be copied and deleted deeply, as well as saved and restored from disk.
The Framatome solution involves the following (see also Section~\ref{sec:segmentdesc}): 
(i) each graph indexes node segment by an integer; 
(ii) segments refer to one another by their integer indexes instead of their pointers. 
Therefore, to migrate such a graph design to \fnew, it is sufficient to be able to implement an array of pointers to segments of different types in \fnew.
However, \fnew does not have a direct construction for an array of pointers (see point~\ref{reqt-fortran-arrays-of-pointers}).
We will keep the same mechanism to maximise the developers' familiarity with the migrated code.
We will therefore have a large array of derived types and references to these derived types will continue to be through indexes in this array.
For this to compile, all derived types will inherit from one abstract type that we call \mycode{segment} (see line 5 in Listing~\ref{lst:f2k-example}).
It declares the common \esope commands that all derived types will have to define (abstract methods, see Listing~\ref{lst:abstractsegment}).

\begin{CodeListing}
\begin{lstlisting}[language=Fortran, caption={Abstract \seg in \fnew},label={lst:abstractsegment}]
   type, abstract:: segment
   contains
       procedure(abstract_segsup), deferred :: segsup
       procedure(abstract_segcop), deferred :: segcop
       procedure(abstract_segmov), deferred :: segmov
       procedure(abstract_segprt), deferred :: segprt
       procedure(abstract_seg_store), deferred :: seg_store
       procedure(abstract_seg_type), deferred :: seg_type
   end type segment
\end{lstlisting}
\end{CodeListing}

\subsection{\esope commands}  \label{sec:migrate-commands}

\esope has a library  of commands to manipulate  \segs (Section~\ref{sec:segcmd}).
For this, \esope could resort to its own meta-description of the segments.
To migrate these commands, we implement them as part of the derived type declaration, inside the module for this derived type.
For example we will have a \mycode{user\_segprt(...)}, or \mycode{user\_segini(...)}.
Since we specialize the command for each segment, we know their structure and can create these commands accordingly without requiring meta-description at compile time.

The aforementioned abstract derived type \mycode{segment} declares the common behavior that all derived types must have.
Here are the details:
\begin{itemize}[wide]
\item The two commands \mycode{segini}, and \mycode{segadj} are not defined in the \mycode{segment} abstract derived type because their parameters include the dimensioning variables and thus cannot be defined generically.
For example, the ``user'' derived type has one dimensioning variable \mycode{ubbcnt}.
In our running example, line 10 in Listing~\ref{lst:esope-example} will become \mycode{call segini(ur, ubbcnt)}.
It is considered a bonus that the two commands will now require as explicit parameter the dimensioning variables of the segment (see Section~\ref{sec:pb-nontech}, point~\ref{reqt-nontech-dimensioninvars});

\item The two commands \mycode{segsup}, and \mycode{segprt} are part of the \mycode{segment} abstract derived type.
They take only one parameter: the instance they manipulate.
They will be used with the statement: \mycode{call segsup(p)};

\item The two commands ``segcop'', and ``segmov'' are introduced.
They do not really exist in \esope but were informal names given by the developers to alternative forms of \mycode{segini} and \mycode{segact} (see Section~\ref{segcmd-segcop}).
They are now declared as parts of the \mycode{segment} abstract derived type.
They take two parameters: the target instance and source instance.
For example, \mycode{call segmov(p, q)} will copy the contents of the derived type pointed by $q$ into the derived type pointed by $p$ (without first allocating memory for $p$);

\item The two commands \mycode{segdes}, and \mycode{segact} that were used to swap memory out or in are removed as swapping will now be handled automatically.
\end{itemize}

\subsection{Additional \esope notations}  \label{sec:migrate-esope-utility}

We have also established translation rules for accesses to \seg fields, as shown in the following examples.
It will follow the standard \fnew syntax:
\begin{itemize}[wide]
  \item \mycode{p.scalar} or \mycode{p.a($i_1$, \dots, $i_n$)} will be respectively translated to \mycode{p\%scalar} and 
  \mycode{p\%a($i_1$, \dots, $i_n$)}.

  \item The \esope slash expression, used to get the size of an array (\mycode{dim = p.a(/1)}) will translate to \mycode{dim = size(p\%a, dim=1)}.
\end{itemize}

\subsection{Preprocessor statements}  \label{sec:migrate-preproc}

File inclusion (\mycode{\#include}) was used in \esope code, often to ``import'' the declaration of a segment (very much like a \mycode{.h} file inclusion in C), and sometimes to reuse the declaration and initialization of a group of variables, or even some statements such as a loop.

All the includes are removed from the migrated code.
Segment declarations are migrated in their own module as explained in Section~\ref{sec:migrate-segment},
every other statement is copied from the included file to where the file was originally included.

Currently the comments in the included files are not copied.
One reason is that there are sometimes several pages of comments that would enlarge too much the including files.
Another difficulty is that it is difficult to assign automatically a comment to the entity it describes.
So, it is difficult to know if a comment would need to be copied in the module of a derived type or in the program unit that includes a file.

Since segments are now modules they need to be imported (\mycode{use} in \fnew) in the program units that need them.
The \mycode{use} statements are computed afterwards, looking at everything that is required but not define by a program unit.

As a special traceability feature (Section~\ref{sec:pb-nontech}, point~\ref{reqt-nontech-traceability}) we generate a comment line to mark the beginning and end of all included code.

Others preprocessor statements such as \mycode{\#define}, \mycode{\#ifdef} will continue to be treated by the preprocessor (\mycode{cpp}).

\subsection{Implicit typing}  \label{sec:migrate-implicit}

Fortran allows one to use variables without declaring them.
In this case, the variable's type is inferred from its name\footnote{Typically, variable names starting with the letters `i' through `n' denote integers, while other names denote reals.}.
An \mycode{implicit none} statement forbids to use the implicit typing.
It is largely present in \esope code, but very often replaced by \mycode{implicit logical (a-z)} just before calling the Esope transpiler,
which does not accept \mycode{implicit none} (see Section~\ref{sec:pb-esope}).
However, not all code use such replacement statements.
As a consequence, some variables were implicitly typed by mistake.

Concurrently, there are also cases where \mycode{implicit none} was not used,
but a specific implicit declaration instead\footnote{For example \mycode{implicit integer (x)}
that states that all variable names starting with `x' are integer instead of the normal real.}.

In the migrated code, we remove all implicit statements and insert a mandatory \mycode{implicit none} in all files.
This may cause errors at compilation for the few variables that were inadvertently not declared explicitly.
We also ensure that all variables used in the code are explicitly declared by inferring their type using the standard Fortran naming rules.
\subsection{Parameter intent}  \label{sec:migrate-intent}

In \fold parameters are passed by reference by default.
\fnew provides an \mycode{intent} declaration to specify:
\begin{itemize}[wide]
  \item \mycode{in}: the parameter's value in the caller is used in the called, but not modified;
  \item \mycode{out}: the parameter does not have an initial value provided by the caller, but is used to return a value;
  \item \mycode{inOut}: the parameter's value from the caller is used in the called, and the (possibly modified) value is returned.  
\end{itemize}

We could have declared all parameters as \mycode{inOut} in \fnew.
But to improve code quality and reduce errors, we aim to determine the correct intent of all the parameters.
This is based on some simple rules:
\begin{itemize}[wide]
\item If the parameter is first accessed before being assigned, then it must be at least \mycode{in};
\item Conversely, if it is first assigned before being accessed, then it must be \mycode{out};
\item If it is first access and then assigned, it is \mycode{inOut} (because it was passed by reference in \fold);
\item Parameters passed along to other routines depend on the intent of these other routines to know whether they are an access or an assignment;
\item We created a catalog of predefined routines with their known parameter intents.
\end{itemize}

In the next section, we will discuss some details of the migration process and tool.

\section{Migration Implementation} \label{sec:migration-implementation}

In this section we explain how we implemented the \esope migration.
The solution implements the choices defined in the previous section.
It is based on Model Driven Engineering (MDE): the \esope source code is represented as a model (an AST); this model is transformed in a \fnew model; and finally, the resulting model is exported into \fnew code.
We also detail the migration process.
The result is applied on an example use case and we evaluate the result.

\subsection{Model Driven Migration}

We chose to implement the migration based on metamodel transformation.
The solution is based on the Famix~\cite{Anqu14a,Anqu20a}
and FAST\footnote{\href{https://github.com/moosetechnology/FAST}{https://github.com/moosetechnology/FAST}}
metamodels~\cite{Anqu20a}.

Famix is a dependency metamodel representing the ``user defined entities'' of a programming language (\eg classes, functions, parameters,\ldots) and the dependencies between these entities (\eg invocations to functions, accesses to variables).
It does not contain enough detail to allow reproducing (or migrating) the source code.
The FAST (Famix AST) metamodel is a complete AST representation of the source code with expression, statements, identifiers,\ldots

The Famix model is more concise: it allows one to model a large quantity of code (millions of lines of code if needed).
It also represents all dependencies across the entire application, for example between entities in different source files.

Containing all the details of the source code, FAST is a larger model.
It would not be possible to have the complete AST of the application in memory.
But this level of detail is required to be able to regenerate the code (in its old form or migrated to \fnew).

Famix offers mechanisms to load and unload on the fly the FAST model of a given subprogram.
Nodes inside the AST (for example a call statement) are mapped to nodes inside Famix (a function invocation).

\subsection{Migration process}
The migration process includes the following steps:
\begin{itemize}[wide]
  \item Step 1: Parse Esope source into a Famix model.
  \item For all included files (that appear in some \mycode{\#include} instruction):
  \begin{itemize}
  \item Step 2: Import the content of the file into a FAST model.
  \item Step 3: For any segment definition encountered, export the generated derived type into \fnew.
  \item Step 4: Any remaining statements are copied to the location of the \mycode{\#include} in the host subprogram.
  \end{itemize}
  \item For each not included file, gather the subprograms it contains and for each of these:
  \begin{itemize}
  \item Step 5: Import the FAST model for that subprogram.
  \item Step 6: Migrate the \esope FAST model to a corresponding \fnew FAST model.
  \item Step 7: Export the \fnew FAST model generated into a source file. 
  \end{itemize}
\end{itemize}

In our approach, we treat included files as program units.
However, we treat them separately from other program units because the segment definitions are migrated to their own module, and other instructions are copied where they are included in other program units. This is explained in step 3 and 4.

\paragraph{Parsing.}
Parsing \esope code (steps 1, 2 and 5) is not trivial as there was no easily re-usable parser for it and the grammar was not formally defined.
We will not go, here, in the details of how this problem was solved.
In a nushell, we used a small island grammar parser that copied the \esope code into \fold code by putting in comments the \esope commands so that a standard \fold parser can parse the code.
From that modified code, we generate a Famix model including the comments.
We then go through this model again to generate Famix \esope entities from the statements in the comments.
This process is explained in more details in \cite{Sow23a}.

\paragraph{Migration.}
The migration itself (steps 3 and 6) is done through model transformation following the MDE approach.
Because we must go through the entire \fold AST to migrate it to a \fnew AST, we chose to use a visitor (instead of transformation rules for example).
We use out-of-place \cite{Kaha19a} transformation with two distinct models:
For a \fold subprogram, we visit its AST from the root (subprogram node) down to the leaves, and for each node, create a replacement in the \fnew AST model.
Pure \fold nodes are usually passed through as they are in the new model because \fnew is backward compatible.
We introduce a module node as root of the migrated program units to contain each subprogram.
\esope nodes are transformed into \fnew subtrees in the resulting AST.
Segment declarations are special in that they result in their own \fnew AST.

In Step 3, the result of migrating pure Fortran code (\textit{ie.} not the segment declarations) of the included file is recorded in a dictionary of all included files.
These partial models are then inserted in-place whenever a \mycode{\#include} for a given included file is encountered.
For readability, we also enclose these inserted nodes between two comment nodes.

Finally, exporting the migrated model to \fnew source code (steps 4 and 7) is straightforward, as the AST contains all the information needed.

\subsection{Generating large chunks of code}

When we migrate a \seg, we face code inflation due to the generation of all \esope commands for each derived type.

Constructing the AST for these commands would have been tedious and error-prone.
It would also make the migration tool difficult to maintain due to the sheer size of the code generating these large ASTs.

To simplify the migration process and make it easier to write and maintain the migration tool, we use plain strings and string templates with placeholders (\eg \mycode{procedure ::} \mycode{segini => \{1\}\_segini}, where ``\{1\}'' is a placeholder).

One challenge is inserting child nodes into a typed AST, where the parent node expects children of a specific type. 
For instance, an assignment statement node expects a variable as its left child and an expression as its right child.
Likewise, a subprogram definition node requires a specific structure comprising a name, parameters and body, each of which has its own expected type.
Inserting strings in this AST cannot be done anywhere.

We can use the string nodes to represent a single statement or a complete program unit. 
To be able to do this, we use the traits of Pharo\footnote{\href{https://pharo.org/}{https://pharo.org/}} (similar to interfaces with default methods in Java) for these strings.
A string node (plain or template) can be both a statement or a subprogram thanks to traits.

These string nodes make it easier to generate a large amount of similar code.
But they also have some downsides:
\begin{itemize}[wide]
	\item Testing the compliance of string nodes with the \fnew meta-model is not possible.
	They are in the model but outside the meta-model;
	
	\item Correct formatting of the code is also more challenging. For example, indentation depends on the nesting level of the instructions, which may vary for various uses of the same string.
\end{itemize}

\subsection{\esope sample migration use case}

We applied the solution to a specifically crafted \esope representative sample called ``Bookstore''.
The Bookstore simulates a library that manages users, books, and loans. 
Three Segments represent the library, the books, and the users.

The Bookstore example was manually crafted before this project started to include all the peculiarities of \esope and evaluate whether \fnew could represent a valid replacement for \esope.
We extended this example to add large arrays that would be more representative of the actual applications to be migrated: Each Book contains a large array of integers (between $100$ and $8\ 000$ elements) and the library is randomly populated: $100$ users, $1\ 000$ books, $100\ 000$ actions (loans or returns).

These are some data on the bookstore use case:
\begin{itemize}[wide]
  \item $23$ files comprising $1611$ lines of \esope code.
  \item $3$ \segs: a user, a book, and an library.
  \item $16$ subroutines, $3$ functions, and $1$ main program.
\end{itemize}

\subsection{Use case migration}

The total migration takes about 6 seconds on a recent laptop.
The resulting source code includes:
\begin{itemize}[wide]
  \item $23$ files comprising $3545$ lines of \fnew code.
  \item $3$ derived types from the $3$ segments.
  \item $22$ modules (from $16$ subroutines, $3$ functions and $3$ segments) and $1$ main program.
\end{itemize}

\paragraph{Behavior evaluation:}
The generated code compiles and runs without error.
The output of the migrated application is the same as for the \esope one.

\paragraph{Speed Evaluation:}
We also performed an evaluation of the performances although the small size of the project gives weak guarantees that the results are representative of real HPC applications. We compiled the \esope and \fnew examples using the same vendor and version of the compiler and the same compilation options.
We use the \mycode{time} Unix command to compare the time performance of executions.

\begin{table}[htbp]
  \centering
  \caption{Execution time (in seconds) for four versions of the bookstore: original \esope; \esope{}* with lower memory allocation; \fnew with invariant checks; \fnew{}* without invariant checks.}
  \label{tab:result-time}
  \begin{tabular}{ll@{ }ll@{ }ll@{ }l}
  \hline
  & \multicolumn{2}{c}{\textbf{Real}} & \multicolumn{2}{c}{\textbf{User}} & \multicolumn{2}{c}{\textbf{Sys.}} \\ 
  & Mean  & Std & Mean & Std & Mean & Std \\
  \hline
  \textbf{Esope}  & 4.627&0.158 & 3.309&0.044 & 1.292&0.147 \\

  \textbf{Esope*}  & 3.025&0.027 & 2.987&0.025 & 0.017&0.006 \\

  \textbf{Fortran 2008} & 1.681&0.017 & 1.658&0.016 & 0.009&0.005 \\

  \textbf{Fortran 2008*} & 1.028&0.026 & 1.007&0.026  & 0.009&0.004 \\
  \hline

  \end{tabular}
\end{table}

Table~\ref{tab:result-time} shows the execution times averaged over 100 executions.
Looking at the first and third rows, one can see that the total time (``Real'')
for migrated \fnew is less than half that of \esope.
\esope spent quite a bit of CPU time in system mode (``Sys'').
The explanation can be memory allocation as \esope starts by requesting a large amount of memory ($2$ gigabytes by default) for its own heap.
We tested this hypothesis by reducing the total amount of heap memory allocated at startup.
We lowered this amount to $32$ megabytes, and the results are shown in the second row (\esope{}*) in 
Table~\ref{tab:result-time}.
It does reduce the \esope system time ($0.017$ sec.) to become closer to \fnew ($0.009$ sec.).
Yet \esope ``User'' time is still much more than \fnew.

\fnew has extra ``user'' time due to some new invariant checks that were added in the migrated code.
The last row (\fnew{}*) shows the time for the migrated version without these checks.
We see that the time is even better for \fnew{}* migrated code.

The final result is that \fnew{}* is about three time faster than \esope{}*.
We stress again, however, that this simple application is not comparable to the real applications of Framatome and this excellent result must not be taken as a definitive evaluation.

\section{Discussion} \label{sec:future-work}

Although the migration is now working, there are still some points to be dealt with.
We discuss here future steps to take and limitations of our approach.

\paragraph{Comments:}
Comments in reverse engineering are traditionally difficult to deal with because they are not part of the language's grammar.
It can be difficult to automatically decide to what entity a comment pertains.
Therefore when migrating entities to different parts of the code, it is difficult to know where the comment should go.
We have the case with segment definition and the declaration of their dimensioning variables.
The first goes in a separate module while the second goes anywhere the segment is used.
For now we do not deal with this issue and all these comments go with the ``segment module'' (module defining the derived type for the segment).

\paragraph{\esope constructs:}
We are not yet dealing with all \esope constructs (\eg \mycode{segini p=q} or \mycode{segact p=q}) and the parser needs to be extended.

\esope code base relies heavily on outdated constructs: \mycode{GOTO}, \mycode{COMMON}, and \mycode{EQUIVALENCE}.
Eliminating them is a relevant and useful research topic \cite{Cecc14a, Dool99a, Morr97a, Seym01a}, but it is outside the scope of this project.
Because we are migrating to \fnew, we do not need to worry right now about these outdated constructs.
We will rely on past results \cite{Dool99a, Orch13a} to treat them.

\paragraph{Validation:}
The use case has some limitations.
The real applications to migrate have been around since the early 1990s, and are the combined work of many developers.
By contrast, the Bookstore  was created more recently by a single developer and may therefore reflect their best practices.

As is often the case, the real source code mixes, \fold conventions with newer ones (Fortran 90 statements).
We have yet to implement a parser that transparently accepts all these conventions and keeps the comments (see~\cite{Sow23a} for details).

\paragraph{Code inflation:}
The migrated \fnew derived type includes the definitions of all the \esope commands for each derived type.
This results in a large code expansion (hundreds of lines of code for a segment) that makes reading them cumbersome and modifying them difficult.

The developers will have to maintain the code in the future and modify the segments' definition.
As expressed in Section \ref{sec:pb-nontech} the definition of the segments should be lightweight, easy to read, understand and modify.

We are considering using a companion DSL (Domain Specific Language) for this as well as a small tool to convert the DSL into \fnew.
The DSL would express only the data structure of the segment just as is now the case with \esope.
The tool would then convert this data structure into the \fnew definition with the ``boilerplate'' code for the \esope commands (Section \ref{sec:migrate-commands}).
In this vision, the developers would never look at the Fortran segments modules but only work with the application code and the DSL segment definition.

Two options are currently under study for the DSL:
(i) a simple textual description like TOML\footnote{\url{https://toml.io/en/}};
(ii) keeping the old \esope format (for which we already have a parser) which is close to Fortran syntax.

\section{Conclusion} \label{sec:conclusion}

\fold, a programming language in use since the $1980$s, remains essential for
many scientific applications where performance is critical.
\esope extends \fold offering additional features: structured type and dynamic memory allocation.
However, the evolution of compilers and development technologies has limited
certain essential optimisations, which could compromise the long-term sustainability
of these applications. This raises the fundamental question wether  \esope applications will still be compilable in the future?

To address this challenge, we developed an approach that automatically migrates \esope applications to modern Fortran (\fnew). 
We presented all the difficulties and requirements of this migration and detailed how they were addressed.

Tests on an \esope example application demonstrated that this migration does not affect the validity of the code, with the same results between the \esope and \fnew versions. Furthermore, our evaluations of execution time and memory usage showed that the migrated application is faster and uses less memory than the \esope one.

This approach paves the way for ensuring the longevity of \esope legacy applications 
while leveraging these applications by using \fnew with modern compilers.
Future work and research will be required to extend this process to more complex cases and
to keep pace with the evolution of compilers. 

\bibliographystyle{plainnat}
\bibliography{
  ./rmod,
  ./others
}

\end{document}